\documentstyle[a4,12pt]{article}
\begin{document}
\large
\begin{center}
{\bf Temporal behaviour of emissions from  $\gamma-$ray bursts and 
optical/near-IR afterglows of GRB 991208 and GRB 991216 }
\end {center}
\normalsize

\centerline {\sc R. Sagar$^1$, V. Mohan$^1$, A. K. Pandey$^1$, S. B. Pandey$^1$ 
and A. J. Castro-Tirado$^2$}
\medskip
\centerline {\it $^1$U. P. State Observatory, Manora Peak, Nainital -- 263 129,
 India}
\centerline {\it $^2$IAA-CSIC, P.O. Box 03004, E-18080, Granada, Spain}
\bigskip

\begin{abstract}
The CCD magnitudes in Cousins $R$ and $I$ photometric passbands are determined 
for GRB 991216 and GRB 991208 afterglows respectively $\sim$ 1 and $\sim$ 3 day 
after trigger of the corresponding $\gamma-$ray bursts. Light curves of 
the afterglow emissions are obtained by combining the published data with the 
present measurements in $R$ and $I$ passbands for GRB 991208 and 
in $R$, Gunn $i$ and $J$ passbands for GRB 991216. They indicate that the flux 
decay constants of a GRB are almost the same in each passband with values 
$\sim 2.2$ for GRB 991208 and  $\sim 1.2$ for GRB 991216 indicating very fast
optical flux decay in the case of former which may be due to beaming effect. 
However, cause of steepening by $0.23\pm0.06$ dex in the $R$ light curve of GRB 
991216 afterglow between 2 to 2.5 day after the burst is presently not 
understood. Redshift determinations indicate that both GRBs are at cosmological 
distance with a value of 4.2 Gpc for GRB 991208 and 6.2 Gpc for GRB 991216. The 
observed fluence $>$ 20 keV indicates, if isotropic, release of energy 
$\sim 1.3\times 10^{53}$ erg for GRB 991208 and $\sim 6.7\times 10^{53}$ erg 
for GRB 991216 by 
these bright $\gamma-$ray flashes. The enormous amount of released energy will 
be reduced, if the radiation is beamed which seems to be case for GRB 991208 
afterglow.  The quasi-simultaneous broad-band photometric spectral energy 
distributions of the afterglows are determined $\sim$ 8.5 day and $\sim$ 35
hour after the bursts of GRB 991208 and GRB 991216 respectively. 
The flux decreases exponentially with frequency. The value of spectral
index in the optical-near IR region is $-0.75\pm$0.03 for GRB 991208 and 
$-1.0\pm$0.12 for GRB 991216. 
\end{abstract}


\section {Introduction}

Gamma-ray bursts (GRBs), electromagnetically the most luminous events in the 
Universe, are short and intense flashes of cosmic high energy ($\sim$ 100 
keV$-$1 MeV) photons. The emission from GRB afterglows, though similar in 
nature to the emission from supernovae, is more energetic by a few orders of
magnitudes, as they release $\sim 10^{51} - 10^{54}$ ergs or more in a few 
seconds. Following the naming sequence, nova and supernova, it is therefore 
appropriate to call GRBs as hypernova. The origin of GRBs is a mystery even 
after about 30 years of their accidental discovery by the Vela satellites. 
Interest focused on where they came from and what they were, but was hampered
by lack of sufficient data to even locate them.  A major break through in our 
understanding of GRBs has been achieved in recent years (cf. Kulkarni et al. 
1999; Galama et al. 1999; Castro-Tirado et al. 1999a and references therein) 
mostly due to multi-wavelength observations of the long-lived emission, known
as afterglow of GRB, at $X-$ray, optical and radio wavelengths 
which have become routine after launch of the Italian-Dutch $X-$ray satellite 
BeppoSAX in mid-1996, as it provides the positions of GRBs with  an accuracy 
better than 3$-$5 arc minutes within hours of occurrence. They indicate that 
most likely all GRBs are at cosmological distances. 

\medskip
	The fireball plus blast wave is the most accepted current theoretical 
model for GRBs and their afterglows (see Piran 1999 for a review). 
The GRBs are thought to arise when a massive explosion, known as a fireball, 
releases a large ($\sim M_{\circ}c^2$) amount of kinetic energy into a volume of 
less than 1 light ms across. When this ultra-relativistic outflow of particles 
interacts with surrounding material, both forward and reverse shocks are formed.
The GRB itself is thought to owe its multi-peaked light curve to a series of 
internal shocks within a relativistic flow while the afterglows are 
due to the external (forward) shocks driven in the interstellar medium 
surrounding the burster. As the external shock interacts with increasing 
amount of swept-up material, it becomes less relativistic, and produces a 
slowly fading afterglow of $X-$ray, then ultra-violet, optical, infrared, 
millimeter and radio radiation. The afterglow emissions are most likely 
synchrotron radiation (Sari et al. 1998; Piran 1999 and references therein). 
Compared to the duration of the GRBs, afterglows in the long-wavelength bands 
can be long-lived. This makes now a days international multi-wavelength 
observing campaigns as integral part of a GRB research. The optical transient 
(OT) of a GRB has generally apparent $R$ 
magnitude between 18 to 22, if it is detected a few hours after the burst. 
The 1-m class optical telescopes equipped with CCD detector are therefore 
capable of detecting the optical early afterglow of a GRB. As the optical 
follow-up observations of the GRB afterglows are valuable for understanding 
the nature of these bursts, we started such observations at U.P. State 
Observatory (UPSO), Nainital in January 1999 using 104-cm telescope and CCD 
detector under an international collaborative programme coordinated by one of 
us (AJCT). So far, successful photometric observations have been carried out 
for 3 GRB afterglows from the UPSO, Nainital. The UPSO photometric 
observations for GRB 990123 have been presented by Sagar et al. (1999). The 
same for the other two, namely GRB 991208 and GRB 991216 are presented here. 
These in combination with data published in GCN Observational reports are used 
to study flux decay and spectral index in optical and near-infrared(IR) 
regions. An introduction to the GRBs studied here is given below.

\subsection{GRB 991208}

Hurley et al. (1999) reported  Ulysses, Russian Gamma-Ray Burst Experiment 
(KONUS) and Near Earth Asteroid Rendezvous (NEAR) detection of an extremely 
intense, 60 s long GRB on 1999 December 8 at 04:36:52 UT with a fluence $>$ 25 
keV of $10^{-4}$ erg cm$^{-2}$ and considerable flux at $>$ 100 keV. 
Observations taken on 1999 December 10.92 UT with VLA at 4.86 GHz and 8.46 GHz 
by Frail et al. (1999) indicate presence of a compact source of spectral index 
of $+1.4$ at 
$\alpha_{2000}=16^h 33^m 53.^s5; \delta_{2000}=+46^{\circ} 27^{'} 21^{''}$ with 
a strong candidacy for the afterglow from GRB 991208. At the same location, 
optical afterglow of GRB 991208 was identified first by Castro-Tirado et al. 
(1999b) and confirmed by optical observations reported by Stecklum et al. (1999)
and Jensen et al. (1999a). Coincident (within errors) with the location of 
optical and radio afterglows, Shepherd et al. (1999) detected at millimeter 
wavelengths the brightest afterglow of a GRB reported so far. At 15 GHz and
240 GHz, the GRB 991208 afterglows have been observed by Pooley (1999a) and 
Bremer et al. (1999) respectively. As this GRB seems to be unusually bright 
at $\gamma-$rays, optical, millimeter and radio wavelengths; its detailed study 
may provide new clues regarding the origin of GRB phenomena. The optical 
spectra of the GRB taken on 1999 December 13 and 14 with the SAO-RAS 6-m 
telescope (Dodonov et al. 1999b) indicate a redshift of $z = 0.7055\pm 0.0005$,
which was later also confirmed by Djorgovski et al. (1999a) with the Keck 
spectra taken on 1999 December 14 and 15. 

\subsection{GRB 991216}

Kippen et al. (1999) reported Burst and Transient Source Experiment (BATSE)
detection of an extremely bright $\gamma-$ray burst on 1999 December 16 at 
16:07:01 UT (trigger No. 7906) with total fluence above 20 keV $\sim 2.6\times 
10^{-4}$ erg/cm$^2$. The event was also detected by NEAR. The burst is thus one 
of the brightest event detected by both BATSE and NEAR with spectral properties 
typical of a GRB. Its optical afterglow was detected by Uglesich et al. (1999) 
at $\alpha_{2000}=05^h 09^m 31.^s29;  \delta_{2000}=+11^{\circ} 17^{'} 07^{''}$.
Coincident with this position, Taylor $\&$ Berger (1999) and  Pooley (1999b) 
detected a radio source at 8.5 GHz and 15 GHz respectively, while Takeshima 
et al. (1999) discovered an $X-$ray afterglow. Djorgovski et al. (1999b) 
detected the host galaxy of the burst at $z$ = 1.02 (Vreeswijk et al. 1999b), 
which extends out to $\sim$ 1$^{''}$ to the west of GRB 991216 afterglow.

\section { Optical observations, data reduction and calibrations } 

The optical observations were carried out from 
1999 December 12 to 14 for the GRB 991208 afterglow and on 1999 December 17 for 
the GRB 991216 afterglow. We used a 2048 $\times$ 2048 pixel$^{2}$ CCD system 
attached at the f/13 Cassegrain focus of the 104-cm Sampurnanand telescope of 
the UPSO, Nainital. As GRB 991208 was located mostly in the 
day light sky, making optical observations was a herculean task. On all three 
nights, observations were obtained just before the morning twilight at large 
air-mass ($>2$) but in good photometric sky conditions. Long durations of 
nights at Nainital in the month of December helped us to carry out 
observations at least for an hour or so on each day. One pixel
of the CCD chip corresponds to 0.$^{''}$38, and the entire chip covers a field
of $\sim 13^{'} \times 13^{'}$ on the sky. Fig. 1 shows the location of the 
GRB 991208 and GRB 991216 afterglows on the CCD images taken from UPSO, 
Nainital. For comparison, images extracted from the Digital Palomar Observatory 
Sky Survey (DSS) are also shown where the absence of a GRB OT is clearly seen. 
Several short exposures upto a maximum of 15 minutes were generally given. In 
order to improve the signal-to-noise ratio of the OT, the data have 
been binned in $2 \times 2$ pixel$^2$ and also all images of a night have been 
stacked after correcting them for bias, non-uniformity in the pixels and cosmic 
ray events. Exposure times for the stacked images of GRB 991208 afterglow in $I$
were 400, 1800 and 3600 s on 1999 December 12, 13 and 14 respectively. The total
exposure time for the stacked image of GRB 991216 afterglow in $R$ is 85 minutes 
on 1999 December 17. As the OTs were generally quite faint on the
stacked images, DAOPHOT profile-fitting technique was used for the magnitude 
determination. On 1999 December 12 and 13 the GRB 991208 afterglow was
sufficiently bright and we could derive its magnitude. However, on 1999 December
14 it had become too faint to be measured on the image. On 1999 December 13, we 
also observed Landolt (1992) standard stars to calibrate
$I$ magnitudes of the GRB 991208 afterglow. The results of UPSO observations 
along with other photometric measurements of GRB 991208 OT are given in Table 1.

\medskip
In the field of GRB 991216, three stars (as identified in Fig. 1) are 
photometrically calibrated in $R$ passband by Dolan et al. (1999) using 
Landoldt (1992) standard stars located in SA 93, SA 97 and SA 98 regions. 
The quoted uncertainty in the zero-point calibration is $\pm$0.03 mag.
The $R$ magnitudes determined by Dolan et al. (1999) agree very well with
an independent measurement carried out later on by Henden et al. (2000). This 
indicates that photometric calibration used in this work is secure.
Present $R$ magnitude is relative to comparison star A (see Fig. 1). This along 
with other photometric measurements of GRB 991216 afterglow used in the present 
analysis are given in Table 1. In order to avoid errors arising due to 
different photometric calibrations, we have used only those $R$ measurements 
published in GCN Observational reports whose magnitudes could be determined 
relative to comparison stars shown in Fig. 1. In Gunn $i$ and $JHK$ filters, 
all published photometric measurements have been used.

\newpage
\noindent {\bf Table 1.}~Photometric observations of the GRB 991208 and 
GRB 991216 afterglows. Total errors in magnitude measurements are mostly
$\geq$0.1 while statistical errors are always $\le$0.1.

\medskip
\begin{tabular}{ccll} \hline 
Time in UT &Filter & Magnitude & Source  \\  \hline 
  \multicolumn{4}{c}{ GRB 991208 afterglow}  \\ 
	Dec 99	10.27&	R&	18.7$\pm$0.1&Jensen et al. (1999a)	\\
	Dec 99	11.27&	R&	19.5$\pm$0.1& Castro-Tirado et al. (1999c)\\
	Dec 99	11.85&	R	&20.0$\pm$0.2&Jensen et al. (1999a)\\
	Dec 99	12.29&	R&	20.25$\pm$0.15& Masetti et al. (1999)	\\
	Dec 99	12.52&	R&20.5$\pm$0.1&Garnavich \& Noriega-Crespo (1999)\\
	Dec 99	13.29&	R&	20.3$\pm$0.2& Masetti et al. (1999)	\\
	Dec 99	13.53&	R&	20.92$\pm$0.06&	Halpern \& Helfand (1999) \\
	Dec 99	14.14&	R&	21.6$\pm$0.3&	Dodonov et al. (1999a) \\
	Dec 99	14.29&	R&	21.25$\pm$0.15& Masetti et al. (1999)	\\
	Dec 99	15.29&	R&	21.6$\pm$0.3& Masetti et al. (1999)	\\ 
	Dec 99	11.21&	I&	19 to 19.5&	Stecklum et al. (1999) \\
	Dec 99	12.02&	I&	19.4$\pm$0.18&	Present work \\
	Dec 99	13.02&	I&	19.9$\pm$0.14&	Present work \\
	Dec 99	14.00&	I&	$>$ 20&	Present	work 	 \\ 
        Dec 99  16.68&  K &     19.31$\pm$0.15 & Bloom et al. (1999) \\ \hline
  \multicolumn{4}{c}{ GRB 991216 afterglow}  \\ \hline
	Dec 99	17.148&	R&18.63$\pm$0.02&Dolan et al. (1999) \\
	Dec 99	17.152&	R&18.64$\pm$0.02&Dolan et al. (1999) \\ 
	Dec 99	17.179&	R&18.73$\pm$0.05&Henden  et al. (1999) \\ 
	Dec 99	17.216&	R&19.00$\pm$0.05&Henden  et al. (1999) \\ 
	Dec 99	17.293&	R&19.06$\pm$0.05&Henden  et al. (1999) \\ 
	Dec 99	17.448&	R&19.25$\pm$0.04&Dolan et al. (1999) \\
	Dec 99	17.455&	R&19.28$\pm$0.04&Dolan et al.  (1999) \\
	Dec 99	17.610&	R&19.48$\pm$0.10&Jha et al. (1999) \\
	Dec 99	17.733&	R&19.89$\pm$0.13&Giveon et al. (1999) \\
	Dec 99	17.780&	R&19.70$\pm$0.08& Present work \\
	Dec 99	18.110&	R&20.12$\pm$0.10&Jensen et al. (1999b) \\
	Dec 99	18.320&	R&20.40$\pm$0.10&Jensen et al. (1999b) \\
	Dec 99	18.320&	R&20.32$\pm$0.05&Garnavich et al. (1999) \\
	Dec 99	18.400&	R&20.30$\pm$0.06&Diercks et al. (1999b) \\
	Dec 99	18.560&	R&20.57$\pm$0.05&Garnavich et al. (1999) \\
	Dec 99	19.100&	R&20.90$\pm$0.10&Jensen et al. (1999b) \\
	Dec 99	29.410&	R&23.60$\pm$0.30&Djorgovski et al. (1999b) \\ 
        Jan 00  06.181&	R&24.21$\pm$0.12&Schaefer (2000) \\
	Dec 99	17.214&Gunn i&19.10$\pm$0.20& Diercks  et al. (1999a) \\ 
	Dec 99	17.462&Gunn i&19.90$\pm$0.20& Diercks  et al. (1999a) \\ 
	Dec 99	17.350&	J&16.99$\pm$0.05&Garnavich et al. (1999) \\
	Dec 99	18.130&	J&17.56$\pm$0.02&Vreeswijk et al. (1999a) \\
	Dec 99	18.300&	J&18.25$\pm$0.06&Garnavich et al. (1999) \\ 
	Dec 99	18.130&	H&16.74$\pm$0.02&Vreeswijk et al. (1999a) \\
	Dec 99	18.130&	K&16.76$\pm$0.02&Vreeswijk et al. (1999a) \\ \hline
\end{tabular}

\section{Prompt $\gamma-$ray emission}

For GRB 991208, prompt $\gamma-$ray emissions were detected by KONUS 
and NEAR while in the case of GRB 991216, they were detected by BATSE and NEAR.
We have downloaded the light curves from the archive and shown them in Figs.
2 and 3 for GRB 991208 and GRB 991216 respectively. Presence of multi-peaked
spiky temporal profile in the energy distributions of both GRBs is an 
unambiguous indicator of a series of internal shocks within a relativistic flow.
Further discussions on the $\gamma-$ray light curves of each GRB are given below.

\subsection {GRB 991208}

Fig. 2 shows the light curve in two energy bands accumulated by KONUS in 50 --
200 keV and by NEAR in 100 -- 1000 keV. The burst profile is dominated by two 
strong peaks, separated by about 55 s. Almost identical temporal as well as
intensity structures in the light curves at both energy bands indicate that
perhaps, dominant emission is in the 100 -- 200 keV range as it is common in 
both observed energy bands. The burst began with a strong pulse which lasted for
$\sim$ 6 s. It has a sharp rise and a relatively slow decline. This is followed by
a relatively weak pulse starting at $\sim$ 40 s after the trigger. It is a 
relatively broad profile. Almost at the end of this pulse and $\sim$ 52 s after
trigger of the burst, the strongest pulse of this GRB started and lasted 
for $\sim$ 20 s. This has almost the same rise and decline time, though the profile 
is multi-peaked, asymmetric and irregular. As expected, the spiky nature of the 
profile is clearly visible only on the 64 ms time resolution light curve.  
Duration (full width at half maxima) of the profile at trigger of the burst is 
only $\sim$ 2 s while that of the strongest one is more than 5 s. They also 
differ in temporal structures. The GRB is a long duration burst as it lasted 
for more than 70 s.

\subsection {GRB 991216}

We show in Fig. 3 the light curves of GRB 991216 in four energy bands, 
obtained by the BATSE on board the Compton Gamma-Ray Observatory satellite. 
The light curve obtained by NEAR is not shown here as it resembles to the 
BATSE highest energy light curve and also has poor time resolution.  The burst 
has a complicated and irregular time profile. The event began with a weak 
precursor pulse lasting about 2 s, followed $\sim$ 15 s later by an intense 
multi-peaked complex. The main emission lasted for $\sim$ 17 s followed by a 
fainter tail that persisted for 
another $\sim$ 20 s (Fig. 3). The $T_{50}$ and $T_{90}$ durations of the burst, 
as measured in the 50 -- 300 keV energy range, are 6.272$\pm$0.09 s and 
15.168$\pm$0.11 s respectively (Kippen 1999) indicating that it is a long 
duration burst. The burst lasted for $\sim$ 60 s and it has peaks of width
$\sim$0.5 s, yielding a value of the variability index as 120.

\medskip
The overall shape of the GRB 991216 in all the
energy bands can be described as a fast rise starting at $\sim$ 17 s; arrived
maximum $\sim$ 21 s and then decayed slowly. Each phase of the burst profile 
contains a number of well-defined short duration (full width at half-maxima 
generally $\le$ 0.5 s) sub-pulse or spikes within the burst. There are 14 such 
spikes. We list in Table 2 their time of occurrence and relative counts with 
respect to the first spike which has 24.3, 38.9, 49.1 and 8.5 K count/s above 
the background in the energy bands 
20 -- 50 keV, 50 -- 100 keV, 100 -- 300 keV and $>$ 300 keV respectively.
Spikes 1, 2 and 3 are during ascending phase; 4, 5 and 6 are during maxima phase
and others are during descending phase of the burst. The peak of spikes occurred
almost simultaneously in all the four energy bands. The variations in relative
count rates of a spike from higher to lower energy bands are similar for spikes
of a phase but differ from the spikes of other phases. The relative count 
rates are nearly the same in all energy bands for the spikes during ascending 
phase; they are more in $>$ 100 keV energy range than in 20 -- 100 keV range 
for spikes around maxima and they decrease systematically with increase in 
energy for spikes of descending phase. The light curves show that the low-energy
emission persists longer and peaks later than high-energy emission. 
Particularly striking is the paucity of $>$ 300 keV emission during the 
shoulder about 32 s after trigger of the burst. In the highest energy band 
emission is maximum in spike 4 and is relatively much reduced at lower energy 
bands. {\bf Hard-to-soft} spectral evolution thus observed in GRB 991216 is 
the typical one usually (but not always) seen throught the GRB and also in
sub-pulses within a burst (Fishman et al. 1999 and references therein).

\vspace{0.5cm}
\noindent {\bf Table 2.}~List of well-defined $\gamma-$ray spikes during GRB 
991216. Time of its occurrence ($T_p$) in 20 -- 50 keV energy band after 
trigger of the burst at 1999 December 16.671 UT along with their relative count 
rate with respect to first spike in all the four $\gamma-$ray energy bands 
are given. Spikes are identified with the ascending, maxima and descending 
parts of the GRB 991216 burst. 

\begin{tabular}{cccccc l} \hline 
Spikes&$T_p$&\multicolumn{4}{c}{Relative count rate in keV energy band}&Phase \\ 
      &	(s) & 20 -- 50 & 50 -- 100 & 100 -- 300 & $>$ 300 \\  \hline 
 1  & 17.3 & 1.000 & 1.000 & 1.000 & 1.000 & Ascending  \\ 
 2  & 19.3 & 1.693 & 1.591 & 1.554 & 1.611  & Ascending   \\ 
 3  & 19.7 & 1.771 & 1.804 & 1.810 & 1.705  & Ascending   \\ 
 4  & 19.9 & 1.775 & 1.809 & 2.197 & 3.526  & Maxima   \\ 
 5  & 20.6 & 2.381 & 2.356 & 2.561 & 2.586  & Maxima    \\ 
 6  & 21.9 & 2.216 & 2.089 & 2.368 & 2.452  & Maxima    \\ 
 7  & 22.4 & 1.610 & 1.360 & 1.212 & 0.847  & Descending    \\ 
 8  & 22.7 & 2.402 & 2.402 & 2.470 & 1.752 & Descending      \\ 
 9  & 23.0 & 2.072 & 1.719 & 1.187 & 0.589 & Descending      \\ 
 10  & 23.7 & 1.313 & 0.978 & 0.601 & 0.307 & Descending      \\ 
 11  & 25.6 & 1.124 & 0.836 & 0.575 & 0.313  & Descending     \\ 
 12  & 26.8 & 1.033 & 0.679 & 0.369 & 0.154 & Descending      \\ 
 13  & 29.5 & 1.887 & 1.614 & 1.291 & 0.918 & Descending      \\ 
 14  & 31.1 & 1.940 & 1.616 & 1.244 & 0.713 & Descending      \\ \hline
\end{tabular}

\medskip
In the energy band 50 -- 300 keV, peak flux at 64, 256 and 1024 ms intervals
are 19.93$\pm$0.24, 17.80$\pm$0.11 and 14.55$\pm$0.13 $\mu$erg/cm$^2$/s 
respectively while the fluence in the BATSE energy channels 1, 2, 3 and 4 are 
17.2637$\pm$0.053, 22.9693$\pm$0.054, 65.0656$\pm$0.136 and 150.204$\pm$1.167 
$\mu$erg/cm$^2$ respectively (Kippen 1999). The hardness ratio 
$\frac{f_{100-300}}{f_{50-100}}$ is thus = 2.83$\pm$0.02.

\section{ Optical and near-IR photometric observations }

Both GRB 991208 and GRB 991216 afterglow emissions have been photometred in 
optical and near-IR passbands and the results are published in GCN 
Observational reports. We have used these data in combination with the present
measurements to study their flux decay and
spectral index from 0.7 $\mu$m to 2.2 $\mu$m in the following sub-sections.

\subsection{ Light curves}

In order to study the optical and near-IR photometric light curves of the 
GRB 991208 and GRB 991216 afterglows, we have plotted photometric measurements 
as a function of time in Fig. 4. The X-axis is log ($t-t_0$) where $t$ is the 
time of observation given
in Table 1 and $t_0$ is the time of the trigger of GRB event which is 1999 
December 8.192 UT for GRB 991208 and 1999 December 16.671 UT for GRB 991216. All
times are measured in unit of day. The decay of earlier GRB afterglows appears 
to be well characterized by a power law $F(t) \propto (t-t_0)^{-\alpha}$, 
where $F(t)$ is the flux of the afterglow at time $t$ and $\alpha$ is the decay
constant.  Assuming this parametric form and by fitting least square linear 
regressions to the observed magnitudes as function of time, we derive below
the value of flux decay constant for both GRB 991208 and GRB 991216 afterglows. 

\subsubsection { GRB 991208}

Present observations in combination with the publised ones have been used to 
derive flux decay constants of the OT of GRB 991208. Fig. 4(A) 
shows the light curve in  $R$ and $I$  photometric passbands where the useful
measurements in $I$ are only from UPSO, Nainital. The object is fading very 
fast, and no flattening of the light curves is observed.  We obtained the 
following linear relations for the $R$ and $I$ magnitudes as function of time \\
 $ R(t) = (16.95\pm 0.06) + (5.47\pm 0.09) log (t-t_0) $ \\
 $ I(t) = (16.52\pm 0.01) + (4.95\pm 0.01) log (t-t_0) $ \\
These lines are also plotted in Fig. 4(A). In $R$, points which are discrepant 
by more than 3 $\sigma$ have been excluded from the analysis. If one includes
them, then the slope becomes steeper with a value of 5.92$\pm$0.55. In $I$, the
linear relation is derived only from the UPSO observations. Other observations 
in $I$ (see Table 1) are not accurate enough to be used for deriving the linear 
relation. Allowing for the factor $-2.5$ involved in converting the flux to 
magnitude scale, the values of $\alpha$ are either $2.15\pm0.04$ or 
2.37$\pm$0.22 in $R$ and $\sim 2 $ in $I$ passbands. This indicates that flux 
decays in $R$ and $I$ are almost similar and hence, we adopt a value of 
2.2$\pm$0.1 as flux decay constant in further discussions. GRB 991208 afterglow 
is thus decaying much faster than afterglows of other GRBs observed so far 
including GRB 990123 afterglow observed earlier by us (cf. Sagar et al. 1999) 
where $\alpha$ is 1.1$\pm$0.06. This is also the brightest 
afterglow detected at millimeter wavelengths to date (Shepherd et al. 1999). 
Thus GRB 991208 OT presents an interesting case for understanding the origin 
of radiation from a $\gamma-$ray burst across the entire electromagnetic band. 

\subsubsection { GRB991216}

In Fig. 4(B), we have plotted photometric measurements in $R$, Gunn $i$ and
$J$ passbands as a function of time. The UPSO, Nainital measurement, identified 
in the figure, fits very well with the observed linear decay of the $R$ 
magnitude. The emission from GRB 991216 OT is fading in all 3 passbands. The 
light curves do not exhibit flattening in any passband, though observations in 
$R$ are until $\sim$ 20 days after the GRB. We obtained the following linear 
relations for the  $R$, $i$ and $J$ magnitudes as function of time \\
 $ R(t) = (19.69\pm 0.02) + (3.04\pm 0.10) log (t-t_0) $ \\
 $ i(t) = (19.91\pm 0.01) + (3.05\pm 0.01) log (t-t_0) $ \\
 $ J(t) = (17.55\pm 0.01) + (3.31\pm 0.01) log (t-t_0) $ \\
These lines are also plotted in Fig. 4(B). In $R$,  16 points respresenting 
early observations ($<$ 2.5 day after the burst) seem to follow a linear 
relation and value of the slope is derived only based on them. In $i$ and $J$ 
passbands, linear relations are derived only from the two observations provided 
by Diercks et al. (1999a) and Garnavich et al. (1999) respectively. The $J$ 
magnitude of  Vreeswijk et al. (1999a) at 1999 December 18.13 UT lies well 
above the linear relation (Fig. 4(B)). By including this point in the linear 
fit, the slope becomes flatter with a value of 2.77$\pm$0.89. 
Allowing for the factor $-2.5$ involved in converting the flux to magnitude 
scale, the values of $\alpha$ are 1.22$\pm$0.04 in $R$, $\sim 1.22$ in $i$ and 
between 1.11$\pm$0.36 to 1.32 in $J$ passbands. A comparison of these indicates 
that at early times ($<$ 2.5 day after the burst), flux decays in optical and 
near-IR regions are almost similar and thus, the values of $\alpha$ are 
independent of wavelength at least in the range of 0.7 to 1.3 $\mu$m. For 
further discussions, we adopt a value of 1.2$\pm$0.1 for the flux decay constant
of GRB 991216. This is in agreement with
the early time decays of most of the GRB afterglows observed so far. 

\medskip
Extrapolation of the $R$ linear relation derived above at the times of $R$ 
measurements of 1999 December 29.41 UT and 2000 January 6.18 UT gives values 
which are brighter than the observed ones at $\sim 2 \sigma$ level. Somewhat 
steeper index is needed to fit these observations. 
Fitting of the least square linear regression to the last three points, shown
as dotted line in the figure, yields $\alpha = 1.45\pm0.04$. This indicates
a steepening of  $\delta \alpha = 0.23\pm0.06$ in the $R$ light curve. More
photometric observations, including those of later dates are required to 
ascertain this observed steepening in the light curve.

\subsection{Spectral index of the GRB 991208 and GRB 991216 afterglows }

The flux distribution of both afterglows has been determined in 
the wavelength range of 0.7 $\mu$m to 2.2 $\mu$m using broadband photometric
measurements listed in Table 1. We used the reddening map provided by Schlegel, 
Finkbeiner \& Davis (1998) for estimating Galactic interstellar extinction 
toward the bursts and found negligible for GRB 991208 but a value of $E(B-V) = 
0.634$ mag for GRB 991216. In converting the magnitudes into fluxes, we have 
used the effective wavelengths and normalisations by Bessell (1979) for $R$ and 
$I$ and by Bessell \& Brett (1988) for $J, H$ and $K$. The fluxes thus derived 
are accurate to $\sim$ 10\%. The Fig. 5 shows the spectrum for both GRB 991208 
and GRB 991216 OTs. We fitted the observed flux distribution with a power law 
$F_{\nu} \propto  \nu^{\beta}$, where $F_{\nu}$ is the flux at frequency $\nu$ 
and $\beta$ is the spectral index. Other details are given below.

\subsubsection {GRB 991208}

We have constructed the GRB 991208 afterglow spectrum on 1999 December 
16.68 UT. This epoch was selected for the long wavelength coverage 
possible at the time of observations of $K$ magnitude. Optical flux at the 
wavelengths of $R$ and $I$ passbands has been derived using the slope of the 
fitted light curve shown in Fig. 4(A) for the present epoch assuming that 
there is no interstellar absorption. The derived fluxes are 4.8, 5.8 and 
12.3 $\mu$Jy at the effective wavelengths of $R, I$ and $K$ passbands 
respectively. They are plotted in Fig. 5(A). It is observed that as the 
frequency decreases the flux increases with $\beta = -0.75\pm 0.03$. This agrees
very well with the value of $\beta = -0.77\pm 0.14$ given by Bloom et al. (1999)
for the spectral index between optical to IR wavelengths. A day earlier on 1999 
December 15.64, the fully calibrated Keck 10-m spectrum give the value of 
$\beta =-0.9\pm 0.15$ between 0.385 $\mu$ to 0.885 $\mu$ 
(Djorgovski et al. 1999a) which agrees with the optical to IR spectral index. 

\subsubsection {GRB 991216}

We have constructed the GRB 991216 afterglow spectrum at about 35 hr after the 
burst, which was selected for the long wavelength 
coverage possible at the time of $JHK$ observations (see Table 1). Using the 
linear equations drawn in Fig. 4(B) and the measurements listed in Table 1, we 
obtained for this epoch $R=20.2, i=20.4, J=17.9, H=16.7$ and $K=16.8$ mag 
with an uncertainty of $0.2$ mag. The Gunn $i$ magnitude has been converted into
Cousins $I$ using $R$ measurement and the relations given by Wade et al. (1979)
and Bessell (1979). This yields $I = 19.5\pm0.2$ mag. These $R, I, J, H$ and 
$K$ magnitudes have been first corrected for Galactic interstellar extinction of
$E(B-V) = 0.634$ mag, which correspond to $A_{Rc}=1.64, A_{Ic}=1.24, A_J=0.56, 
A_H = 0.35$ and $A_K = 0.21$ mag for the standard reddening curve given by 
Mathis (1990) and then converted to fluxes. The fluxes thus obtained are 115, 
130, 190, 305 and 150 $\mu$Jy at the effective wavelengths of $R, I, J, H$ and 
$K$ passbands respectively. The 
results are plotted in Fig. 5(B) for both the observed and the dereddened 
magnitudes, in order to demonstrate the effect of Galactic extinction on the 
shape of the energy distribution of the afterglow. This shows that reddened 
spectrum becomes steeper. It is observed that as the frequency decreases the 
flux increases upto $H$ passband and then it turns over. The spectrum thus may
not be described by a single power law. We find that between $R$ and $H$ the 
spectral slope is $-1.01\pm 0.12$. By ignoring the turnover,
a spectral slope of $-0.45\pm0.30$ is found between $R$ and $K$.

\medskip
The continuum flux of the GRB 991216 afterglow determined at 1999 December 
18.372 UT using $J$ band spectrophotometry by Joyce et al. (1999) rises from 
$\sim$ 150 $\mu$Jy at 1.12 $\mu$m to $\sim$ 220 $\mu$Jy at 1.23 $\mu$m, and 
remains approximately constant at $\sim$ 220 $\mu$Jy out to 1.33 $\mu$m. It is 
thus not particularly well described by a single power law in agreement with 
our findings. This suggest that either not all of the near-IR flux seen is due 
to synchrotron emission, or a break in the synchrotron spectrum was near 
1.25 $\mu$m at the time of observation.

\section{ The energetics of the GRBs}

Redshift determination of $z = 0.7055$ (Dodonov et al. 1999b) for the GRB 
991208 afterglow and of $z = 1.02$ for the host of GRB 991216 (Djorgovski et al.
1999b), yields a minimum corresponding luminosity distances  of 4.2 Gpc
and 6.2 Gpc respectively for a standard Friedmann cosmological model with Hubble
constant $H_0$ = 65 km/s/Mpc, cosmological density parameter $\Omega_0$ = 0.2 
and cosmological constant $\Lambda_0$ = 0 (if $\Lambda_0 > 0$ then the inferred
distances would increase). Considering isotropic energy emission and observed 
fluence above 25 keV of $10^{-4} erg/cm^2$ (Hurley 1999) for GRB 991208 
and observed fluence $>$ 20 keV of 255.5 $\mu$erg/cm$^2$ (Kippen 1999) for GRB 
991216 and using the corresponding inferred luminosity distances, we estimate 
the $\gamma-$ray energy release to be 
$1.3 \times 10^{53}$ erg $\sim 0.07 M_{\circ}c^2$ for GRB 991208 and 
6.7$\times 10^{53}$ erg $\approx$ 0.4 $M_{\circ}c^2$ for GRB 991216.

\medskip
Of the dozen GRBs with known redshifts, five with total fluence energies 
$>$ 20 keV in excess of 10$^{53}$ erg (assuming isotropic emission) are 
GRB 991216 and GRB 991208 (discussed here); GRB 990510 (Harisson et al. 1999); 
GRB 990123 (Andersen et al. 1999; Galama et al. 1999) and  GRB 971214 (Kulkarni 
et al. 1998). The name hypernova, coined recently, to describe such energetic 
events is therefore seems to be most appropriate. Recent observations suggest 
that GRBs are associated with stellar deaths, and not with quasars or the 
nuclei of galaxies as some GRBs including the present GRB 991216 (see 
Djorgovski et al. 1999b) are found off-set from their host galaxy. However, 
release of huge amount of isotropic energy of $\sim 10^{53}$ erg or more is 
essentially incompatible with the popular stellar death models
(coalescence of neutron stars and death of massive stars). The large energy 
release can also not be understood with exotic models such as that of baryon 
decay (Perna \& Loeb 1998), as the entire or a good fraction of rest mass energy
of neutron stars is released in such models. There are two ways namely 
gravitational lensing and non-isotropic emission to reduce the enormous energy 
release. The possibility of the burst being amplified by gravitational lensing 
has been discussed at length by Kulkarni et al. (1999) in the case of GRB 
990123. However, direct observational evidence for lensing lacks in GRB 990123 
and others (cf. Marani et al. 1999). On the other hand, evidence for beaming 
seems to be present in the case of a few GRBs including GRB 991208 discussed 
below. Indeed, almost all energetic 
sources in astrophysics such as pulsars, quasars and accreting stellar black 
holes display jet-like geometry and hence, non-isotropic emission. Beaming 
reduces the estimated energy by a factor of 10 - 300, depending upon the size 
of its opening angle (Sari et al. 1999). The $\gamma-$ray energy released then 
becomes $\leq 10^{52}$ erg, a value within reach of current popular models for 
the origin of GRBs (see Piran 1999 and references therein).

\section{Discussions  and Conclusions }

Using optical and near-IR observations (see Table 1), we obtained the values of 
flux decay constant and spectral index as 2.2$\pm$0.1 and $-0.75\pm0.03$ for 
GRB 991208 OT and 1.2$\pm$0.1 and $-1.01\pm0.12$ for the early time flux decay 
of GRB 991216 afterglows. These values indicate that flux decay of the GRB 
991216 is normal but that of GRB 991208 is fastest in the GRBs observed so far. 
The GRB 991216 OT light curve in $R$ becomes steeper by $\delta \alpha = 
0.23\pm0.06$ at late time ($>$ 2 to 2.5 day after the burst). Before deriving 
any conclusion from the flux decays of these GRBs, we compare them with other 
well studied GRBs. Except GRB 990123 and GRB 990510, all exhibit, at both early 
and late times a single power-law decay, generally $\sim$ 1.2, a value 
reasonable for spherical expansion in the standard model. However, rapid decays 
and flat spectra ($\beta \geq -1.2$) in optical and near-IR region of the OTs of
GRB 980326 with $\alpha$=1.7$\pm$0.13  and $\beta=-0.66\pm0.7$ 
(Castro-Tirado \& Gorosabel 1999, Sari et al. 1999); GRB 980519 with 
$\alpha$ = 2.05$\pm$0.04 and $\beta = -1.05\pm0.10$ (Halpern et al. 1999) and 
GRB 991208 with $\alpha$ = 2.2$\pm$0.1 and $\beta = -0.75\pm0.03$ (discussed 
here), can not be understood in terms of standard models but can either be 
explained by a jet (Sari et al. 1999) or by a circumstellar wind model 
(Chevalier \& Li 1999).

\medskip
Recent theoretical models on beaming of the relativistic outflow 
predict a break and a marked steepening in the afterglow light curve
(M\'{e}sz\'{a}ros \& Rees 1999; Rhoads 1999; Sari et al. 1999). In such models,
a break in the light curve is expected when the jet makes the transition to
sideways expansion after the relativistic Lorentz factor drops below the inverse
of the opening angle of the initial beam. Slightly later, the jet begins a 
lateral expansion which causes a further steepening of the light curve.
The time of occurrence of break and extent of steepening in the 
afterglow light curve thus depend upon the opening angle of the transient 
collimated outflow. At late times, when the evolution is dominated by the
spreading of the jet, the value of $\alpha$ is expected to approach the electron
energy distribution index with values between 2.0 and 2.5 while the value of
$\beta$ is expected to be $-\alpha/2$, if cooling frequency has passed the 
observed frequency and $-\frac{(\alpha-1)}{2}$ otherwise (Sari et al. 1999). 
The expected values of $\alpha$ and $\beta$ are thus in excellent agreement with
the corresponding observed values of $2.4\pm0.02$ and $-0.61\pm0.12$ for GRB 
990510 (Stanek et al. 1999) and 2.2$\pm$0.1 and $-0.75\pm0.03$ for GRB 991208 
(discussed here). Fast optical flux decay and flat broadband spectrum observed 
in GRB 991208 here may therefore be due to effects of beaming like in GRB 
990510. 

\medskip
Observational evidence for a break was found first in the optical light curve 
of GRB 990123 afterglow (Castro-Tirado et al. 1999a; Kulkarni et al. 1999) and
recently in that of GRB 990510 afterglow (Stanek et al. 1999), where the light 
curve steepened 1.5 to 2 days after the burst. The value of 
$\alpha = 1.13\pm0.02$ for early time (3 hr to 2 day) of the GRB 990123 light 
curve  becomes 1.75$\pm$0.11 at late times (2-20 day) while the corresponding 
slopes for GRB 990510 are 0.76$\pm$0.01 and 2.40$\pm$0.02 respectively with the 
break time 1.57$\pm$0.03 day. If the steepening observed in both cases are due 
to beaming, then one may conclude that it occures within $<$ 2 days of the 
burst. We therefore argue that observed steep decay in the optical light curve 
of GRB 991208 afterglow may be due to beaming which occurred before the start 
of its optical observations which is $\sim$ 2.1 day after the burst. However, 
a break in the light curve of GRB afterglow can occur due to a number of causes 
other than beaming (Kulkarni et al. 1999; Wei \& Lu 1999). We are presently 
unable to understand the possible cause of steepening by 0.23$\pm$0.06 in the 
$R$ light curve of GRB 991216 between 2 to 2.5 day after the burst, as beaming 
steepens the light curve by $\sim$ 1 dex. 

\medskip
 Prompt $\gamma-$ray emission light curves of both long duration bursts 
GRB 991208 and GRB 991216 show complicated and irregular time profile. 
Presence of multi-peaked and spiky temporal profile in the light curves of 
both GRBs is an unambiguous indicator of a series of internal shocks within a 
relativistic flow.

\medskip
Redshift determinations yield a minimum distance of  4.2 Gpc for GRB 991208 and 
6.2 Gpc for the GRB 991216, if one assumes standard Friedmann cosmology with 
$H_o = 65$ km/s/Mpc, $\Omega_0 = 0.2$ and $\Lambda_0 = 0$. Both GRBs are thus
at cosmological distances. Considering 
isotropic energy emission, we estimate enormous amount of the $\gamma-$ray 
energy release ($1.3 \times 10^{53}$ erg for GRB 991208 and  
$6.7 \times 10^{53}$ erg for GRB 991216).  These high energetics are reduced
if the emission is not isotropic but collimated, as suggested by the flat 
spectrum and steep decay in the optical light curve of GRB 991208. 
	
\medskip
The quasi-simultaneous spectral energy distributions are determined for both
GRB afterglows in optical and near-IR region. The flux decreases with frequency
in both cases and follows an exponential relation with not too different spectral
slopes. There are indications that the synchrotron spectrum of GRB 991216 afterglow
breaks near 1.2 $\mu$m.

\bigskip
\noindent {\bf Acknowledgements} The authors are thankful to Prof. D. 
Bhattacharya for providing useful comments and suggestions.
This research has made use of data obtained through
the High Energy Astrophysics Science Archive Research Center Online Service,
provided by the NASA/Goddard Space Flight Center.

\medskip
\noindent {\bf References:} 
\begin{itemize}
\item []Andersen M. I.  et al., 1999a, Science, {\bf 283}, 2075
\item [] Bessell M.S., 1979, PASP, {\bf 91}, 589
\item [] Bessell M.S., Brett J.M., 1988, PASP, {\bf 100}, 1134
\item [] Bloom J.S. et al., 1999, GCN Observational Report No. 480
\item[] Bremer M. et al., 1999, GCN Observational Report No. 459
\item [] Castro-Tirado A. J., Gorosabel J., 1999, A\&AS, {\bf 138}, 449
\item [] Castro-Tirado A. J.  et al., 1999a, Science, {\bf 283}, 2069
\item[] Castro-Tirado A. J. et al., 1999b, GCN Observational Report No. 452
\item[] Castro-Tirado A. J. et al., 1999c, IAU Circular No. 7332
\item [] Chevalier R.A., Li Z., 1999, ApJ, {\bf 520}, L29
\item[] Diercks A., Ferrarese L., Bloom J. S., 1999a, GCN Observational Report No.  477
\item[] Diercks A. et al., 1999b, GCN Observational Report No.  497
\item [] Djorgovski S.G. et al., 1999a, GCN Observational Report No. 481
\item [] Djorgovski S.G. et al., 1999b, GCN Observational Report No. 510 
\item [] Dolan C. et al., 1999, GCN Observational Report No. 486
\item[] Dodonov S. et al., 1999a, GCN Observational Report No. 461 
\item[] Dodonov S. et al., 1999b, GCN Observational Report No. 475 
\item [] Fishman J., 1999, A\&AS, {\bf 138}, 395
\item[] Frail D. A. et al.,  1999, GCN Observational Report No. 451
\item [] Galama T.J. et al., 1999, Nature, {\bf 398}, 394
\item[]Garnavich P., Noriega-Crespo A., 1999, GCN Observational Report No. 456
\item[]Garnavich P. et al., 1999, GCN Observational Report No. 495
\item[] Giveon U., Bilenko B., Ofek E., Lipkin Y., 1999, GCN Observational Report No. 499
\item[]Halpern J.P. et al., 1999, ApJ, {\bf 517}, L105
\item[]Halpern J.P., Helfand D.J., 1999, GCN Observational Report No. 458
\item[]Harrison F.A. et al., 1999, ApJ, {\bf 523}, L121
\item[] Henden A. et al., 1999, GCN Observational Report No. 473
\item []Henden A., Guetter H., Vrba F., 2000, GCN Observational Report No. 518
\item[] Hurley K. et al., 1999, GCN Observational Report No. 450
\item[] Jensen B. L. et al., 1999a, GCN Observational Report No. 454
\item[] Jensen B. L. et al., 1999b, GCN Observational Report No. 498
\item [] Jha S. et al., 1999, GCN Observational Report No. 476
\item[] Joyce D., Rhoads J., Ali B., Dell'Antonio I., Jannuzi B., 1999, GCN Observational Report No. 511
\item [] Kippen R. M., 1999, GCN Observational Report No. 504
\item [] Kippen R. M., Preece R. D., Giblin T., 1999, 
         GCN Observational Report No. 463
\item [] Kulkarni S.R. et al., 1999, Nature, {\bf 398}, 389
\item [] Kulkarni S.R. et al., 1998, Nature, {\bf 393}, 35
\item [] Landolt A.U., 1992, AJ, {\bf 104}, 340
\item [] Marani G.F. et al., 1999, ApJ, {\bf 512}, L13
\item [] Masetti N. et al., 1999, GCN Observational Report No. 462
\item [] Mathis J.S., 1990, ARAA, {\bf 28,} 37
\item [] M\'{e}sz\'{a}ros P., Rees M. J., 1999, MNRAS, {\bf 306}, L39
\item [] Perna R., Loeb A., 1998, ApJ, {\bf 509}, L85
\item [] Piran T., 1999, Physics Reports {\bf 314},  575 
\item [] Pooley G., 1999a,  GCN Observational Report No. 457
\item [] Pooley G., 1999b,  GCN Observational report No. 489
\item [] Rhoads J.E., 1999, ApJ, {\bf 525}, 737
\item [] Sagar R., Pandey A.K, Mohan V., Yadav R.K.S., Nilakshi, Bhattacharya 
D., Castro-Tirado A.J.,  1999, BASI, {\bf 27}, 3
\item [] Sari R., Piran T., Halpern J. P., 1999, ApJ, {\bf 519}, L17
\item [] Sari R., Piran T., Narayan R., 1998, ApJ, {\bf 497}, L17
\item [] Schaefer B., 2000, GCN Observational Report No. 517
\item [] Schlegel D.J., Finkbeiner D.P., Davis M., 1998, ApJ, {\bf 500}, 525
\item[]  Shepherd S. et al., 1999, GCN Observational Report No. 455
\item [] Stanek K. Z. et al., 1999, ApJ {\bf 522}, L39
\item[]  Stecklum B. et al., 1999, GCN Observational Report No. 453
\item [] Takeshima T., Markwardt C., Marshall F., Giblin T.,
                     Kippen R. M., 1999, GCN Observational Report No.  478
\item[] Taylor G. B.,  Berger E., 1999, GCN Observational Report No. 483
\item[] Uglesich R.,  Mirabal N., Halpern J., Kassin S., Novati S., 1999,   
                 GCN Observational Report No.  472
\item[]	Vreeswijk P. M. et al., 1999a, GCN Observational Report No. 492
\item[] Vreeswijk P. M. et al., 1999b, GCN Observational Report No. 496
\item []Wade R. A., Hoessel J.G., Elias J.H., Huchra J.P., 1979, PASP, {\bf 91,}
 35 
\item[] Wei D. M., Lu T., 1999, preprint astr-ph/9908273
\end{itemize}
\bigskip
\centerline {\bf Caption to Figure}
\bigskip
\noindent {\bf Figure 1.} 
Finding charts are produced from the CCD images taken from UPSO, Nainital. 
North is top and East is left. For GRB 991208 field, the image is in $I$ 
passband taken on 1999 December 13.0 UT while for GRB 991216, 
it is on 1999 Dec 17.78 UT in $R$ passband. The optical transient (OT) is
located inside the circle. Here only 2.$^{'}4 \times 2.^{'}$4 field of view is 
presented. The region corresponding to CCD image is extracted from the Digital 
Palomar Observatory Sky Survey and marked as DSS. A comparision of both images 
of the same field shows the absence of GRB afterglows on the DSS images. In the 
case of GRB 991208, also shown are the three comparison stars A and B (Jha et 
al. 1999) and C (Henden et al. 1999) calibrated by Dolan et al. (1999) in $R$ 
band.  

\medskip
\noindent {\bf Figure 2.} 
The NEAR and KONUS light curves of GRB 991208 in two energy ranges
50 -- 200 keV and 100 -- 1000 keV. 

\medskip
\noindent {\bf Figure 3.} 
 The BATSE light curves of GRB991216 in four energy ranges 20 -- 50 keV; 
50 -- 100 keV; 100 -- 300 keV and $>$ 300 keV. 

\medskip
\noindent {\bf Figure 4.} 
Light curve of (A) GRB 991208 afterglow in optical $R$ and $I$ photometric
  passbands from 2 to 4 days after the burst while that 
 of (B) GRB991216 afterglow is from 1 to 20 days after 
 trigger of the burst in $R$, Gunn $i$ and $J$ photometric passbands. 
For both GRBs, measurements from UPSO, Nainital have been marked.  

\medskip
\noindent {\bf Figure 5.} The spectral flux distribution of the afterglows of 
(A) GRB 991208 and (B) GRB 991216 at $\sim$ 8.5 day and $\sim$ 35 hr after 
the corresponding bursts. For GRB 991216 both reddened (crosses) and 
unreddened (filled circles) spectra are presented.
	
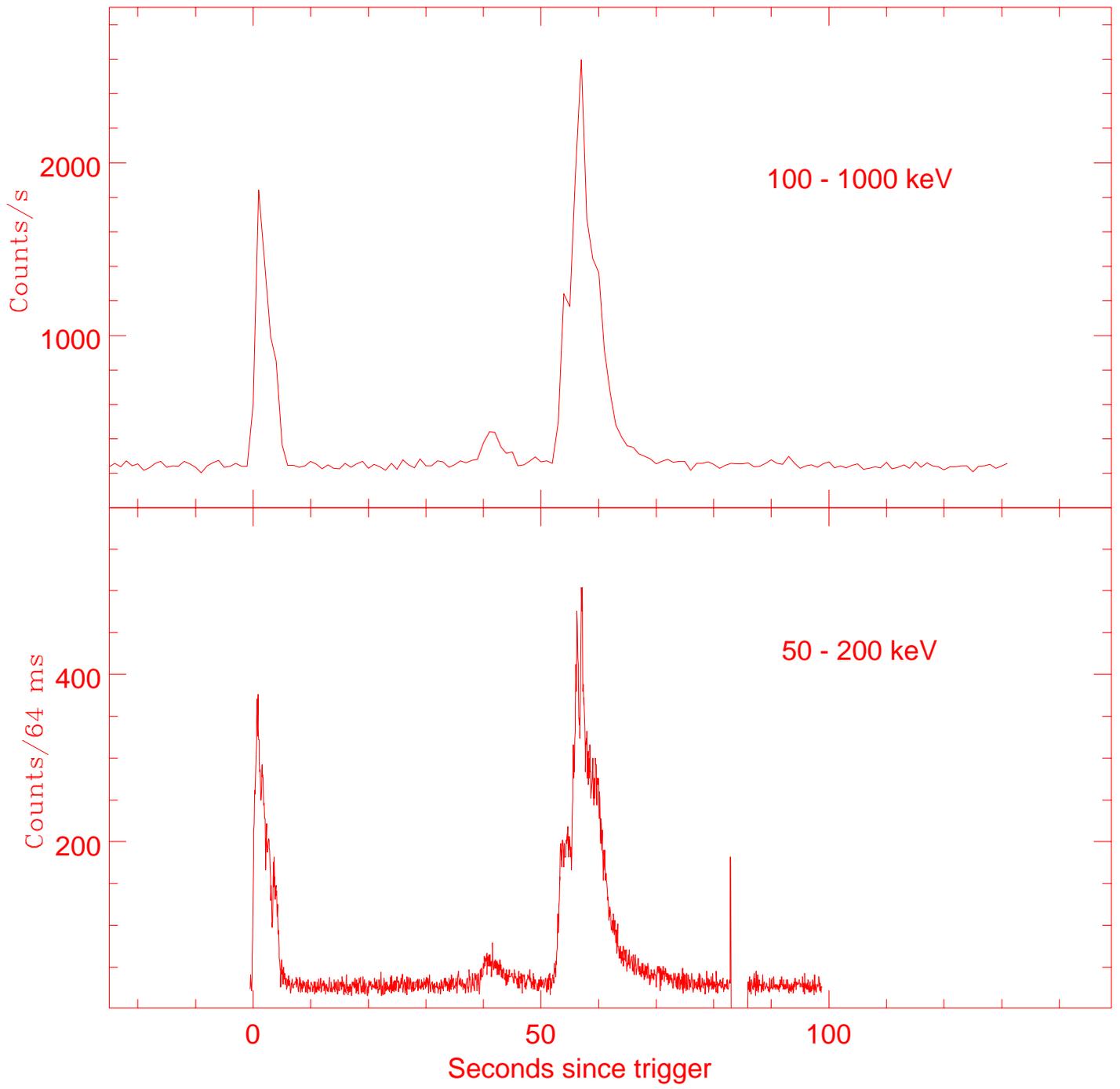
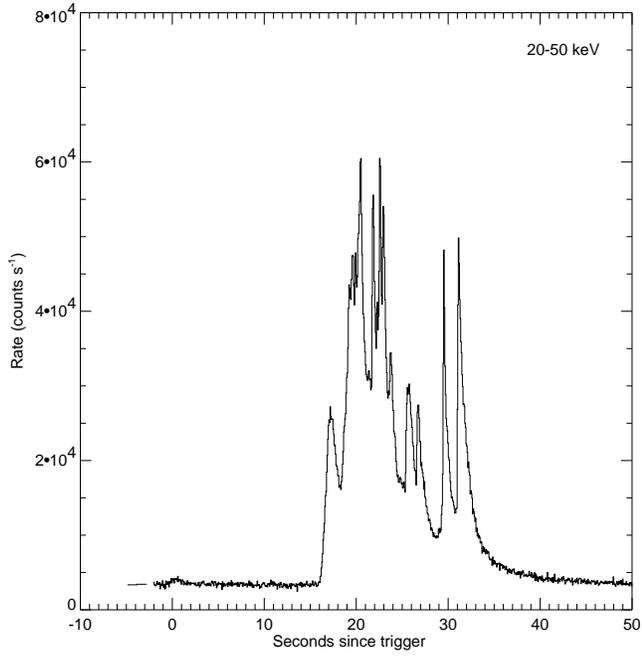
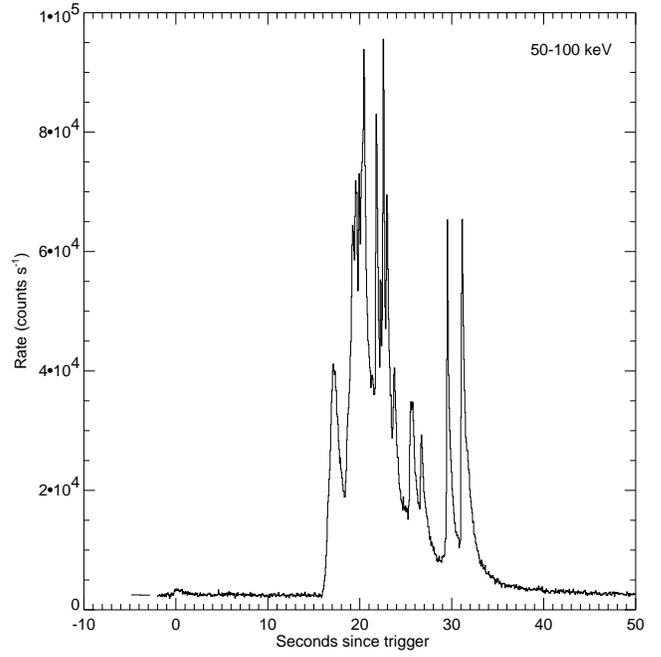
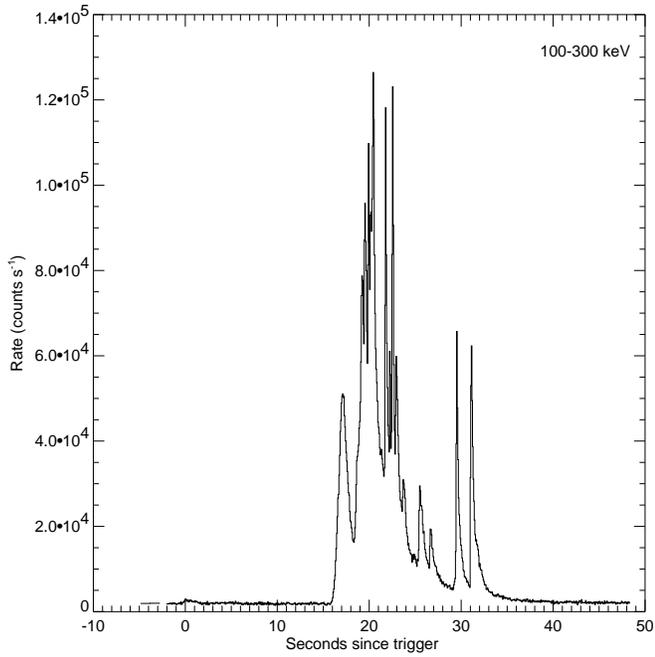
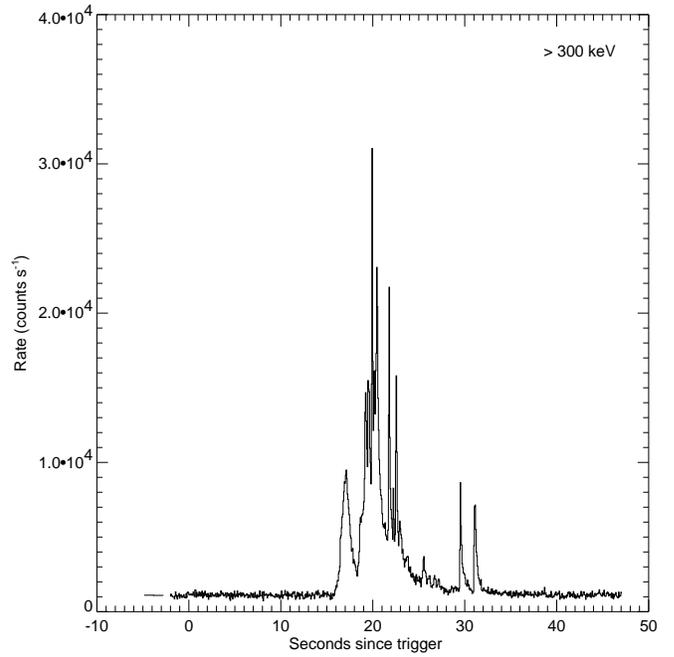
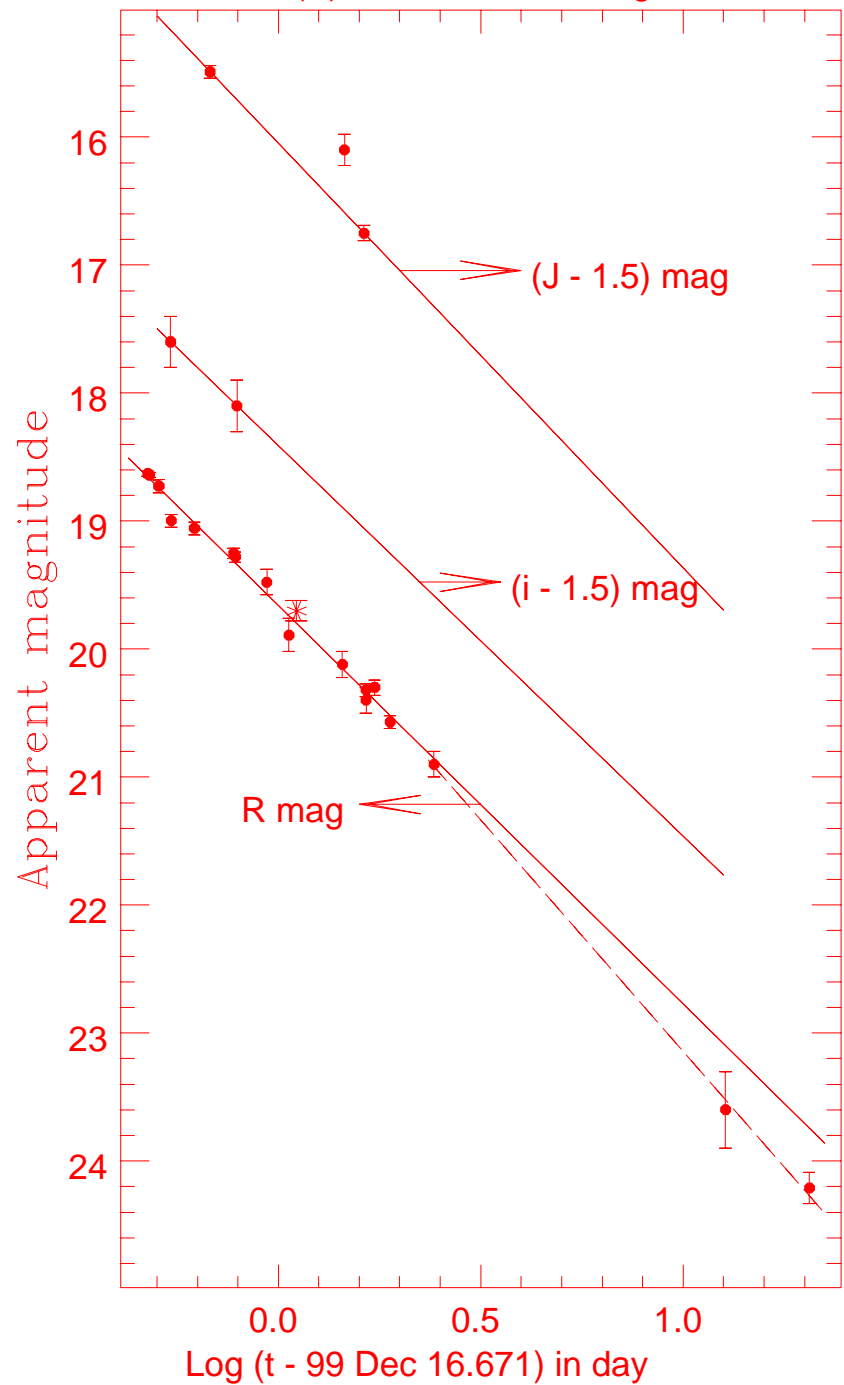
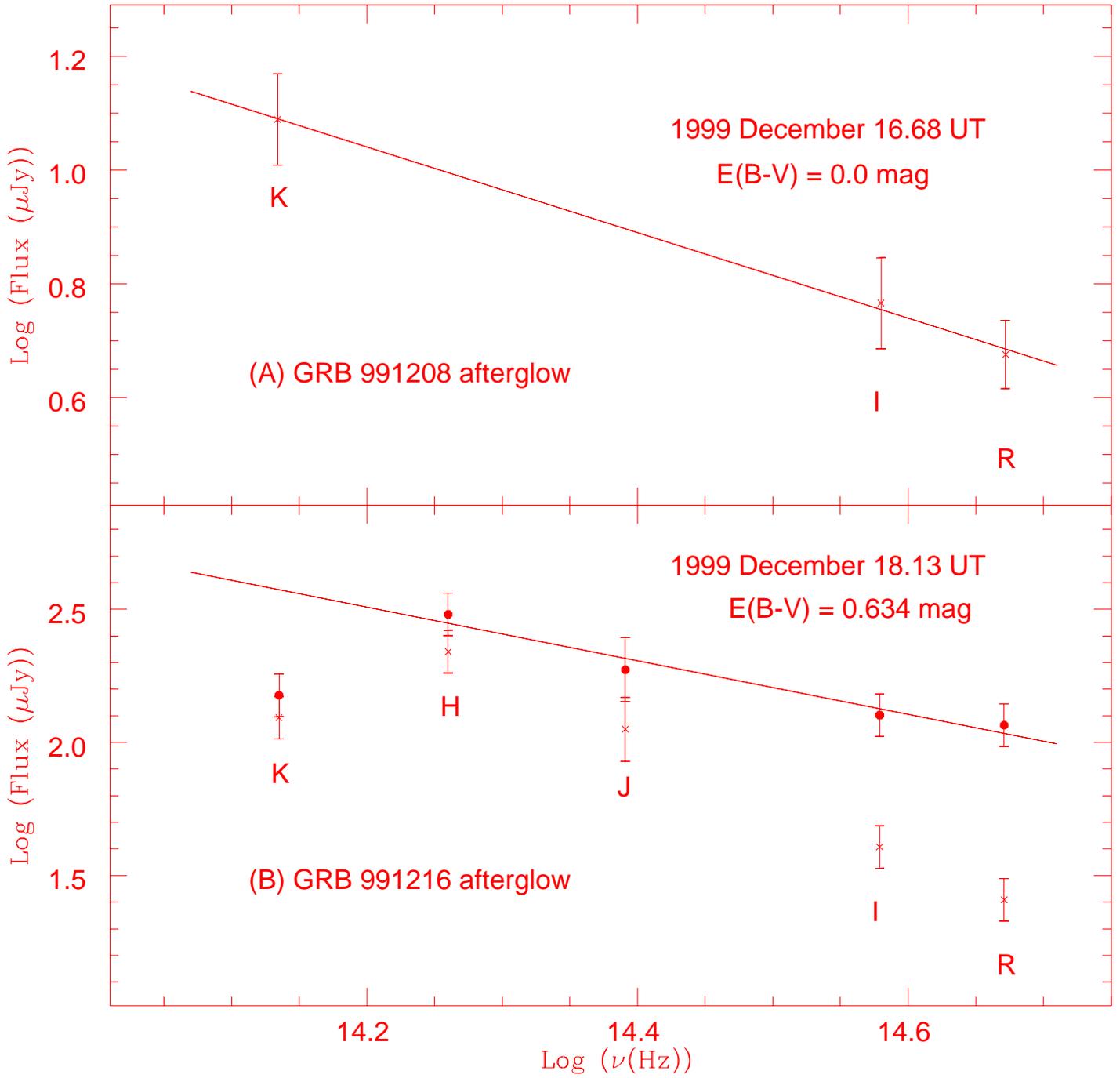
\end{document}